\documentclass[12pt]{article}
\usepackage[dvips]{graphicx}
\usepackage{latexsym,cite}
\def\slashchar#1{\setbox0=\hbox{$#1$}\dimen0=\wd0%
\setbox1=\hbox{/}\dimen1=\wd1%
\ifdim\dimen0>\dimen1%
\rlap{\hbox to
\dimen0{\hfil/\hfil}}#1\else
\rlap{\hbox to \dimen1{\hfil$#1$\hfil}}/\fi}

\newcommand{\beq}{\begin{equation}}
\newcommand{\eeq}{\end{equation}}
\newcommand{\Frac}[2]{\frac{\displaystyle #1}{\displaystyle #2}}

\begin{document}
\thispagestyle{empty}
\begin{titlepage}
\begin{center}
\hfill IFIC/03$-$25 \\ 
\hfill FTUV/03$-$0618 \\
\vspace*{3.5cm} 
\begin{Large}
{\bf Odd--intrinsic--parity processes }\\
{\bf within the Resonance Effective Theory of QCD} \\[2.25cm]
\end{Large}
{ \sc P. D. Ruiz-Femen\'\i a}, { \sc A. Pich} and { \sc J. Portol\'es}
\\[0.5cm]
{\it Departament de F\'\i sica Te\`orica, IFIC, Universitat de Val\`encia -
CSIC\\
 Apt. Correus 22085, E-46071 Val\`encia, Spain }\\[2.0cm]

\begin{abstract}
\noindent
We analyse the most general 
odd-intrinsic-parity effective Lagrangian of QCD valid for processes
involving one pseudoscalar with vector mesons described in terms of 
antisymmetric tensor fields.
Substantial information on the odd-intrinsic-parity couplings is obtained by
constructing the vector-vector-pseudoscalar Green's
three-point function, at leading order in $1/N_C$,
and demanding that its short-distance 
behaviour matches the corresponding OPE result. 
The QCD constraints thus enforced allow us to
predict the decay amplitude $\omega\to \pi \gamma$,
and the ${\cal O}(p^6)$ corrections to 
$\pi\to\gamma\gamma$. Noteworthy consequences concerning the 
vector meson dominance assumption in the decay $\omega\to 3\pi$ are also
extracted from the previous analysis. 
 
\end{abstract}
\end{center}
\vfill
\hspace*{0.5cm} PACS~: 12.38.Aw, 11.40.-q, 12.39.Fe, 13.20.-v, 13.25.-k  \\
\hspace*{0.5cm} Keywords~: Effective theory, QCD, Vector Meson Dominance,
Meson decays. \\ 
\eject
\end{titlepage}

\pagenumbering{arabic}

\section{Introduction}
\hspace*{0.5cm}
Effective field theories of QCD have provided efficient ways to explore
hadron dynamics in those regimes where we are not able to solve the full theory.
Built from the same principles and symmetries which govern QCD, the
effective actions put at our disposal a model-independent framework to 
generate the interactions between the active degrees of freedom. In the very
low-energy domain, chiral perturbation theory
\cite{WE79,GH84,GH85}
has achieved a remarkable success in describing the
strong interactions among pseudoscalar mesons. Moving up to the 1 GeV region
has been proved more difficult, as the effects of vector resonances become
dominant and must be accommodated in the theory. Several works 
\cite{GH84,CO69,EG89}
have provided a sound procedure to include resonance states within the chiral 
framework, later
christened Resonance Chiral Theory. 
This approach, however, leaves the couplings entering the effective Lagrangian
unknown, as they
are not fixed by the symmetry alone. One should then rely on the phenomenology 
or, alternatively, construct theoretical tools that could provide a
meaningful way to compare the results of the effective theory with
those of QCD. The pioneering work of Ref.~\cite{EG89a} indicated that 
the analysis of Green's functions and form factors of QCD currents
yields valuable information on the 
resonance sector and, at the same time, clarifies the ambiguities 
related to the choice of the Lorentz group representation for the 
resonance fields. 
\par
Recently, several authors have pushed forward this direction, either
by using a Lagrangian with explicit resonance degrees of freedom 
\cite{KN01},
or within the framework of the lowest meson dominance (LMD) approximation to
the large number of colours ($N_C$) limit of QCD 
\cite{KN01,BG03,M95,M97,KP99}. In
particular, the authors of Ref.~\cite{KN01} undertook a systematic study of 
several QCD three-point
functions which share the property of being zero in absence of spontaneous 
chiral symmetry breaking for massless quarks.  This common feature 
means that these
Green's functions are free of perturbative contributions from QCD at 
short distances. Therefore, their OPE   
expansion, although formally applicable in the high-energy region, should be
more reliable when descending to energies close to the resonance region, thus
supporting the idea that a smooth matching between QCD and the effective
description involving resonances may exist for these functions. Under
this hypothesis, it was shown in Ref.~\cite{KN01} that while the ansatz derived
from the LMD approach automatically incorporates the right short-distance
behaviour of QCD by construction, the same Green's functions as calculated
with a resonance Lagrangian, in the vector-field representation, are 
incompatible with the OPE outcome. Thus, the  ${\cal O}(p^6)$ 
low-energy constants they extract from the resonance Lagrangian differ
from the estimates of the LMD ansatz. Moreover the authors put forward that
these discrepancies cannot be repaired just by introducing 
local counterterms from the chiral Lagrangian ${\cal L}_{\chi}^{(6)}$, as 
it was done at ${\cal O}(p^4)$ in Ref.~\cite{EG89a}. 
New terms with resonance fields
and higher-order derivatives need to be added, at least in the vector-field
representation, but the general procedure remains unknown.
\par
The result above severely questions the usefulness of the resonance effective
theory beyond the initial work of Ref.~\cite{EG89a}, that rely
not only on the QCD global symmetries but also on the fact that its 
large--$N_C$ limit resembles, at least qualitatively, the three colour theory
\cite{NColor1,NColor2}. In addition one of the basic tenets, after 
the conclusions of Ref.~\cite{NColor2}, is that meson physics in the 
large--$N_C$ limit is described by the tree diagrams of an effective local 
Lagrangian, with local vertices and local meson fields. Hence after the
qualm put forward by Ref.~\cite{KN01} we think that this issue deserves
further investigation. With this aim, we have reanalysed one
of the Green's function studied in this last reference, 
the vector-vector-pseudoscalar
three-point function, this time with the vector mesons described
in terms of antisymmetric tensor fields. 
\par
The latter study requires the introduction
of an odd-intrinsic-parity effective Lagrangian in the formulation of 
Ref.~\cite{EG89} containing
all allowed interactions between two vector objects (currents or resonances)
and one pseudoscalar meson. After a brief introduction on chiral theory, 
Section \ref{sec:RChPT} of this paper is devoted to this subject. 
In Section \ref{sec:short} we evaluate the vector-vector-pseudoscalar 
three-point function $\langle\mathrm{VVP}\rangle$ within our effective
theory at leading order in the $1/N_C$ expansion. We recall its
short-distance properties, as obtained from the OPE calculation, and then we
demand that the $\langle\mathrm{VVP}\rangle$ Green's function built with the
effective action with unknown parameters matches the same behaviour.
The set of relations among couplings derived is then tested in several 
intrinsic-parity-violating decays in Section \ref{sec:pheno}. Finally, we give
our conclusions.

\section{Resonance Chiral Theory and the odd-intrinsic-parity sector}
\label{sec:RChPT}
\hspace*{0.5cm}
The low-energy behaviour of QCD for the light quark sector ($u,d,s$)
is known to be ruled 
by the spontaneous breaking of chiral symmetry giving rise to
the lightest hadron degrees of freedom, 
identified with the octet of pseudoscalar mesons. The corresponding
effective realization of QCD describing the interaction between the
Goldstone fields is called chiral perturbation theory
\cite{WE79,GH84,GH85}.
The effective Lagrangian to lowest order in derivatives, 
${\cal O}(p^2)$, is given by~: 
\begin{equation}
{\cal L}_{\chi}^{(2)}=\frac{F^2}{4}\langle u_{\mu}
u^{\mu} + \chi _+ \rangle \ ,
\label{eq:op2}
\end{equation}
where
\begin{eqnarray}
u_{\mu} & = & i [ u^{\dagger}(\partial_{\mu}-i r_{\mu})u-
u(\partial_{\mu}-i \ell_{\mu})u^{\dagger} ] \ , \nonumber \\ 
\chi_{\pm} & = & u^{\dagger}\chi u^{\dagger}\pm u\chi^{\dagger} u\ \ 
\ \ , \ \ \ \ 
\chi=2B_0(s+ip) \; \; .
\end{eqnarray}
The unitary matrix in flavour space 
\begin{equation}
u(\phi)=\exp \left\{ i\frac{\Phi}{\sqrt{2}\,F} \right\} \; \; \; ,
\end{equation}
is a (non-linear) parameterization of the Goldstone octet of fields~:
\begin{equation}
\Phi (x) \equiv {\vec{\lambda}\over\sqrt 2} \, \vec{\phi}
 = \, 
\pmatrix{{1\over\sqrt 2}\pi^0 \, + 
\, {1\over\sqrt 6}\eta_8
 & \pi^+ & K^+ \cr
\pi^- & - {1\over\sqrt 2}\pi^0 \, + \, {1\over\sqrt 6}\eta_8   
 & K^0 \cr K^- & \bar{K}^0 & - {2 \over\sqrt 6}\eta_8 }.
\label{eq:phi_matrix}
\end{equation}
The external hermitian matrix fields $r_{\mu}$, $\ell_{\mu}$, $s$ and $p$ 
promote the global 
SU$(3)_{\mathrm{R}}\times$SU$(3)_{\mathrm{L}}$ symmetry 
of the Lagrangian to a local one, and generate Green functions of quark currents
by taking appropriate functional derivatives. Interactions with
electroweak bosons can be accommodated through the vector 
$v_{\mu}=(r_{\mu}+\ell_{\mu})/2$ and axial-vector
$a_{\mu}=(r_{\mu}-\ell_{\mu})/2$ fields, while the
scalar field $s$ provides a very convenient way of
incorporating explicit chiral symmetry breaking through the quark masses
$$
s={\cal M}+\dots \ \ \ \ \ , \ \ \ \ {\cal M}=diag(m_u,m_d,m_s) \; \; .
$$ 
The generating functional $Z[v,a,s,p]$ calculated in terms of the external
 sources
is manifestly chiral invariant, but the physically interesting Green functions 
(with broken chiral symmetry) are obtained by taking a particular direction
in flavour space through functional differentiation. Finally,
the ${\cal L}_{\chi}^{(2)}$ Lagrangian is settled by fixing the unknown
$F$ and $B_0$ parameters from the phenomenology~: 
$F \simeq F_{\pi} \simeq 92.4 \, \mbox{MeV}$
is the decay constant of the charged pion and 
$B_0 F^2 = - \langle 0 | \bar{\psi}\psi | 0 \rangle_0$ in the chiral limit.    
\par
Spectroscopy reveals the existence of vector meson resonances as we approach the
1 GeV energy region. These can be classified in SU$(3)_{\mathrm{V}}$ octets and 
must be included as explicit degrees of freedom in order to describe 
hadron dynamics
\cite{CO69}.
At the lowest order in derivatives, the chiral invariant
Lagrangian for the vector mesons and their interaction with Goldstone fields 
reads \cite{EG89}, in the antisymmetric tensor formulation, 
\begin{equation}
{\cal L}_{\mathrm{V}} \, =\,                  
{\cal L}_{\mathrm{Kin}}(V) 
+ {\cal L}_2(V)          
\, ,                                                                            
\label{eq:res_Lagrangian}                                                                            
\end{equation}
with kinetic terms
\begin{equation}
{\cal L}_{\mathrm{Kin}}(V)  =  
    - {1\over 2}\, \langle \nabla^\lambda V_{\lambda\mu}                      
\nabla_{\nu} V^{\nu\mu} -{M^2_V\over 2} \, V_{\mu\nu} V^{\mu\nu}\rangle
\, , 
\label{eq:kin_s}                                                                     
\end{equation}                                                                            
where $M_V$ is the mass of the lowest octet of vector resonances
under SU$(3)_{\mathrm{V}}$, and 
the covariant derivative 
$$
\nabla_{\mu}V=\partial_{\mu}V+[\Gamma_{\mu},V]\ \ \ \ , \ \ \ \ 
\Gamma_{\mu}=\frac{1}{2}\{ u^{\dagger}(\partial_{\mu} - i r_{\mu})u+
u(\partial_{\mu} - i \ell_{\mu})u^{\dagger}\, \} \;  , 
$$
is defined in such a way that $\nabla_{\mu}V$ also transforms as an octet under
the action of the group.                
For the interaction Lagrangian ${\cal L}_2(V)$ we have                
\begin{equation}                                                                           
{\cal L}_2(V)  =   {F_V\over 2\sqrt{2}} \,
\langle V_{\mu\nu} f_+^{\mu\nu}\rangle +                                   
{iG_V\over \sqrt{2}} \, \langle V_{\mu\nu} u^\mu u^\nu\rangle \ ,
\label{eq:V_int}                                                                                                                                                     
\end{equation}
$$
f_{\pm}^{\mu\nu}=u F_L^{\mu\nu}u^{\dagger}\pm u^{\dagger} F_R^{\mu\nu} u \ ,
$$   
with $F_{L,R}^{\mu\nu}$ the field strength tensors of the left and right
external sources $\ell_{\mu}$ and $r_{\mu}$, and $F_V$, $G_V$ are real 
couplings. 
The octet fields are written in the usual matrix notation                       
\begin{equation}                                                                         
V_{\mu\nu} = {\vec{\lambda}\over\sqrt{2}}\,\vec{V}_{\mu\nu}  =            
\pmatrix{                                                       
{1\over\sqrt{2}}\rho^0 + {1\over \sqrt{6}}\omega_{8} 
& \rho^+ & K^{*+}
\cr                                                                 
\rho^- & - {1\over\sqrt{2}}\rho^0 + 
{1\over \sqrt{6}}\omega_{8} & K^{\, *0}
\cr
K^{*-} & \bar{K}^{*0} 
& -{2\over \sqrt{6}}\omega_{8}}_{\!\mu\nu} 
\ .                                                
\label{eq:vmultiplet}                                                                             
\end{equation}                                                                             
\par
The chiral couplings contained in ${\cal L}_2(V)$ 
only concern the even--intrinsic--parity sector. 
In Ref.~\cite{EG89a} it was shown that, up to ${\cal O}(p^4)$ in the
chiral counting, the effective Lagrangian
${\cal L}_{\chi V}\equiv{\cal L}_{\chi}^{(2)}+{\cal L}_{\mathrm{V}}$ 
is enough to satisfy the
short-distance QCD constraints where vector resonances play a significant
role. 
For the odd--intrinsic--parity sector, three different 
sources might be considered~: 
(i) the
Wess-Zumino action \cite{WZ71}, which is ${\cal O}(p^4)$ and fulfills 
the chiral anomaly,
(ii) chiral invariant $\epsilon_{\mu\nu\rho\sigma}$ terms 
involving vector mesons which,  upon integration, 
will start to contribute at ${\cal O}(p^6)$ in the antisymmetric formulation,
and (iii) the relevant operators in the ${\cal O}(p^6)$ Goldstone
chiral Lagrangian \cite{anomchiral}. All of them may contribute to the
$\langle\mathrm{VVP}\rangle$ Green's function. 
\par
The chiral anomaly is driven by the Wess-Zumino action $Z_{\mathrm{WZ}}[v,a]$.
We do not recall its functional here and address
the reader to Ref.~\cite{WZBEG} for the explicit expression. 
On the other side
effective odd-intrinsic-parity Lagrangians with vector resonances have
been previously
considered in the literature in order to study the equivalence of 
different vector resonance models
to reproduce the one-loop divergences of the Wess-Zumino action
\cite{PP93}, 
in the context of the extended Nambu-Jona-Lasinio model
\cite{PR94},
or to estimate the low-energy constants
of the ${\cal O}(p^6)$ Goldstone chiral Lagrangian \cite{KN01}.
Within the antisymmetric formalism, we shall build an independent set of 
odd-intrinsic-parity operators which comprise all possible 
vertices involving two vector resonances and one pseudoscalar (VVP), and 
vertices with one vector resonance and one external vector source plus one 
pseudoscalar (VJP).
\par  
The building blocks for these terms are the ones defined above, which
share the right properties under chiral transformations.
Besides, the terms must satisfy Lorentz, $P$ and $C$ 
invariance. Other useful relations to reduce the number of independent terms
and construct the basis are detailed in the Appendix. Our basis 
reads~\footnote{We
use the convention $\epsilon_{0123}=+1$ for the Levi-Civita tensor 
$\epsilon_{\mu\nu\rho\sigma}$ throughout this paper}:

\vspace{0.5cm}
VJP terms
\begin{eqnarray}
{\cal O}_{\mbox{\tiny VJP}}^1 & = & \epsilon_{\mu\nu\rho\sigma}\,
\langle \, \{V^{\mu\nu},f_{+}^{\rho\alpha}\} \nabla_{\alpha}u^{\sigma}\,\rangle
\; \; , \nonumber\\[3mm]
{\cal O}_{\mbox{\tiny VJP}}^2 & = & \epsilon_{\mu\nu\rho\sigma}\,
\langle \, \{V^{\mu\alpha},f_{+}^{\rho\sigma}\} \nabla_{\alpha}u^{\nu}\,\rangle
\; \; , \nonumber\\[3mm]
{\cal O}_{\mbox{\tiny VJP}}^3 & = & i\,\epsilon_{\mu\nu\rho\sigma}\,
\langle \, \{V^{\mu\nu},f_{+}^{\rho\sigma}\}\, \chi_{-}\,\rangle
\; \; , \nonumber\\[3mm]
{\cal O}_{\mbox{\tiny VJP}}^4 & = & i\,\epsilon_{\mu\nu\rho\sigma}\,
\langle \, V^{\mu\nu}\,[\,f_{-}^{\rho\sigma}, \chi_{+}]\,\rangle
\; \; , \nonumber\\[3mm]
{\cal O}_{\mbox{\tiny VJP}}^5 & = & \epsilon_{\mu\nu\rho\sigma}\,
\langle \, \{\nabla_{\alpha}V^{\mu\nu},f_{+}^{\rho\alpha}\} u^{\sigma}\,\rangle
\; \; ,\nonumber\\[3mm]
{\cal O}_{\mbox{\tiny VJP}}^6 & = & \epsilon_{\mu\nu\rho\sigma}\,
\langle \, \{\nabla_{\alpha}V^{\mu\alpha},f_{+}^{\rho\sigma}\} u^{\nu}\,\rangle
\; \; , \nonumber\\[3mm]
{\cal O}_{\mbox{\tiny VJP}}^7 & = & \epsilon_{\mu\nu\rho\sigma}\,
\langle \, \{\nabla^{\sigma}V^{\mu\nu},f_{+}^{\rho\alpha}\} u_{\alpha}\,\rangle
\;\; ,
\label{eq:VJP}
\end{eqnarray}

VVP terms
\begin{eqnarray}
{\cal O}_{\mbox{\tiny VVP}}^1 & = & \epsilon_{\mu\nu\rho\sigma}\,
\langle \, \{V^{\mu\nu},V^{\rho\alpha}\} \nabla_{\alpha}u^{\sigma}\,\rangle
\; \; , \nonumber\\[3mm]
{\cal O}_{\mbox{\tiny VVP}}^2 & = & i\,\epsilon_{\mu\nu\rho\sigma}\,
\langle \, \{V^{\mu\nu},V^{\rho\sigma}\}\, \chi_{-}\,\rangle
\; \; , \nonumber\\[3mm]
{\cal O}_{\mbox{\tiny VVP}}^3 & = & \epsilon_{\mu\nu\rho\sigma}\,
\langle \, \{\nabla_{\alpha}V^{\mu\nu},V^{\rho\alpha}\} u^{\sigma}\,\rangle
\; \; , \nonumber\\[3mm]
{\cal O}_{\mbox{\tiny VVP}}^4 & = & \epsilon_{\mu\nu\rho\sigma}\,
\langle \, \{\nabla^{\sigma}V^{\mu\nu},V^{\rho\alpha}\} u_{\alpha}\,\rangle
\; \; . 
\label{eq:VVP}
\end{eqnarray}                       
The operators with $\chi_{\pm}$
break SU(3)$_{\mathrm{V}}$ symmetry when distinct quark masses are introduced
through the external scalar field $s=\cal{M}+\dots$. However, only
the pseudoscalar source $p$ in ${\cal O}_{\mbox{\tiny VJP}}^3$ and 
${\cal O}_{\mbox{\tiny VVP}}^2$ will enter our calculation of the 
Green's function, while
${\cal O}_{\mbox{\tiny VJP}}^4$ will not contribute at all and has just been
included in the VJP basis for completeness.
\par
The authors
of Ref.~\cite{PP93} also built the VVP operators in the 
tensor-field representation 
and further constrained the number of independent operators to three by 
applying the equation of motion of the pseudoscalar field at lowest order; 
some care is needed in our case, as particles inside
Green's functions are not on their mass shell. The
resonance Lagrangian for the odd--intrinsic--parity sector will 
thus be defined as
\begin{eqnarray}
{\cal L}_V^{\mathrm{odd}} =&&\!\!\!\!\!\!
{\cal L}_{\mbox{\tiny VJP}}+{\cal L}_{\mbox{\tiny VVP}} \; \; ,
\nonumber\\[3mm]
{\cal L}_{\mbox{\tiny VJP}}=
\sum_{a=1}^{7} \,\frac{c_a}{M_{V}}\,{\cal O}^a_{\mbox{\tiny{VJP}}} 
\ ,   && 
{\cal L}_{\mbox{\tiny VVP}}=
\sum_{a=1}^{4} \,d_a\,{\cal O}^a_{\mbox{\tiny{VVP}}}
\ .
\label{eq:Lano}
\end{eqnarray}
The octet mass $M_{V}$ has been introduced in ${\cal L}_{\mathrm{VJP}}$
to define dimensionless $c_a$ couplings.
We stress that the set defined above is a complete basis for constructing 
vertices with only one-pseudoscalar; for a larger number of pseudoscalars
additional operators may emerge. 
\par
Finally we have to pay attention to the ${\cal O}(p^6)$ Goldstone chiral
Lagrangian. Two operators may contribute at leading order in the 
$1/N_C$ expansion to the  
$\langle\mathrm{VVP}\rangle$ Green's function~: 
\begin{equation}
{\cal L}_{\mathrm{odd}}^{(6)} \, = \, i \, \epsilon_{\mu\nu\alpha\beta}
\left\{ \,t_1\,\langle \,\chi_-f_+^{\mu\nu}f_+^{\alpha\beta}\,\rangle
-i \, t_2 \, \langle \nabla_{\lambda}\,f_+^{\lambda\mu}
\{f_+^{\alpha\beta},u^{\nu}\}\,\rangle \, \right\} \; \; .
\label{eq:WZaction6}
\end{equation}
The $t_i$ couplings are in principle unknown. These operators 
belong both to the effective
theory where resonances are still active degrees of freedom and to the
theory where those have been integrated out. Hence in the latter
case the couplings can
be split as $t_i = t_i^R + \hat{t}_i$
where $t_i^R$ is generated by the integration of resonances and $\hat{t}_i$ is 
a remainder that may survive in the effective theory where resonances are still
active. Vector and pseudoscalar resonances can contribute, in 
principle, to $t_1^R$, though the latter are suppressed because of their
higher masses. Therefore we will consider that 
$t_1^R \simeq t_1^V$. Meanwhile $t_2$ has only vector resonance
contributions and then $t_2^R = t_2^V$. 
Indeed by integrating out the vector mesons in 
${\cal L}_{\mathrm{V}} + {\cal L}_{\mathrm{V}}^{\mathrm{odd}}$
we obtain~:
\begin{eqnarray}
t_1^V&=&-\frac{F_V}{4\sqrt{2}M_V^3}
\left[c_1+c_2+8c_3-c_5\right]+\frac{F_V^2}{8M_V^4}
\left[\,d_1+8d_2-d_3\,\right] \,,
\nonumber\\[3mm]
t_2^V&=&-\frac{F_V}{\sqrt{2}M_V^3}\,(c_5-c_6)+\frac{F_V^2}{2M_V^4}\,d_3
\, .
\label{eq:t1t2}
\end{eqnarray}
On the other side the successful 
resonance saturation of the chiral Lagrangian couplings at ${\cal O}(p^4)$
\cite{EG89} might translate naturally to ${\cal O}(p^6)$ couplings too,
implying that $\hat{t}_i$ could be neglected. We will attach to this 
point and will assume that the $t_i$ couplings are generated completely through
integration of vector resonances. Accordingly we should not include
${\cal L}_{\mathrm{odd}}^{(6)}$ in our evaluation of the Green's function
in order not to double count degrees of freedom. We shall come back to
this discussion in the next Section.
\par
In summary we will proceed in the following by considering the relevant 
effective resonance theory (ERT) given by~:
\begin{equation}
Z_{\mathrm{ERT}}[v,a,s,p] \; = \; Z_{\mathrm{WZ}}[v,a] \, + \, 
Z_{\mathrm{V\chi}}^{\mathrm{odd}}[v,a,s,p] \; \; ,
\label{eq:allZ}
\end{equation}
where $Z_{\mathrm{V\chi}}^{\mathrm{odd}}[v,a,s,p]$ 
is generated by ${\cal L}_{\chi}^{(2)}$ in 
Eq.~(\ref{eq:op2}), ${\cal L}_{\mathrm{V}}$ in Eq.~(\ref{eq:res_Lagrangian})
and ${\cal L}_V^{\mathrm{odd}}$ in Eq.~(\ref{eq:Lano}).

\section{Short-distance information on the odd-intrinsic-parity couplings}
\label{sec:short}
\hspace*{0.5cm} 
The construction of an effective field theory that satisfies the symmetry
requirements of QCD is a model-independent procedure
to accomplish the low-energy properties of the theory 
without missing essential dynamics. 
The price to pay for the universality of 
such approach is an increasing number of (a priori) unknown low-energy 
constants as we tend to improve the accuracy of our calculations, which
eventually reflects in a loss of predictive power. Comparison with data
has been a fruitful way to extract the values of most of the 
chiral couplings up to ${\cal O}(p^4)$, 
as well as some of the resonance parameters for the lightest vector octet
and, to a small extent, for the axial--vector, scalar and pseudoscalar
resonances.  
\par
Jointly with the experimental determination, alternative ways to 
infer the values of the resonance couplings have been explored.
Thus the QCD ruled short--distance behaviour of the vector and axial
form factors in the large-$N_C$ limit (approximated with only one
octet of vector resonances) constrains the couplings of ${\cal L}_2(V)$
in Eq.~(\ref{eq:V_int}), that must satisfy \cite{EG89a}~:
\begin{eqnarray}
1- \frac{F_{V}\,G_{V}}{F^2}&=&0 \; \; \; , 
\label{eq:FVGV}\\[3mm]
2F_{V}\,G_{V}-F_{V}^2&=&0 \; \; \; ,
\label{eq:FVGVa}
\end{eqnarray}
and predict $F_V = \sqrt{2} \, F$ and $G_V = F / \sqrt{2}$, in excellent
agreement with the phenomenology. The strict large-$N_C$ limit would demand 
that the full spectrum of infinite zero--width 
vector resonances should be included in the 
evaluation of the form factors above. However the agreement with data
suggests that 
the approximation of the lightest vector multiplet resembles the limit.
This is the basic assumption of the LMD approach.
\par
In addition, the study of the short-distance properties of
Green's functions and the comparison with the same objects built
from the effective action with explicit resonance degrees of freedom
can yield relevant information on the
resonance couplings, as explored in previous
works \cite{M95,M97,KN01,BG03,KP99}. We now follow this method to 
impose restrictions on the new couplings introduced in the 
odd-intrinsic-parity sector. 
\par
The relevant Green's function for this purpose is the
vector-vector-pseudoscalar QCD 
three-point function $\langle\mathrm{VVP}\rangle$,
\begin{equation}
(\Pi_{\mathrm{VVP}})^{(abc)}_{\mu\nu}(p,q)=\int  d^4x\int d^4y
\,e^{i(p\cdot x+q\cdot y)}
\langle 0|\,T\,[\,V^a_{\mu}(x)\,V^b_{\nu}(y)\,P^c(0)\,]\, |0\rangle \ ,
\label{eq:full3point}
\end{equation}
which requires the octet vector current,
\begin{equation}
V^a_{\mu}(x)=\Big( \bar{\psi}\gamma_{\mu}\frac{\lambda^a}{2}\psi\Big)(x)\,,
\end{equation}
and the octet pseudoscalar density
\begin{equation}
P^a(x)=\Big( \bar{\psi}i\gamma_{5}\frac{\lambda^a}{2}\psi\Big)(x)\,.
\end{equation}

The invariances of QCD under parity and time-reversal
transformations allow us to extract the group and tensor structure of 
$\langle\mathrm{VVP}\rangle$ in the SU$(3)_{\mathrm{V}}$ limit,
\begin{equation}
(\Pi_{\mathrm{VVP}})^{(abc)}_{\mu\nu}=\epsilon_{\mu\nu\alpha\beta}\,
p^{\alpha}q^{\beta}d^{abc}\,\Pi_{\mathrm{VVP}}(p^2,q^2,r^2)\,,
\label{eq:3point}
\end{equation}
with the four-vector $r=-(p+q)$. The first situation 
concerning the short--distance behaviour of the 
$\langle\mathrm{VVP}\rangle$ Green's function that we can analyse is the
case when both momenta $p,q$ in $\Pi_{\mathrm{VVP}}$ become simultaneously
large. The QCD calculation within the OPE framework gives, in the 
chiral limit and up to corrections
of ${\cal O}(\alpha_s)$, \cite{M95}:
\begin{equation}
\lim_{\lambda \to \infty} \Pi_{\mathrm{VVP}}((\lambda p)^2,(\lambda q)^2,
(\lambda p+\lambda q)^2)=
-\frac{\langle \bar{\psi}\psi \rangle_0 }{2\lambda^4}\,
\frac{p^2+q^2+r^2}{p^2q^2r^2}+{\cal O}\left(\frac{1}{\lambda^6}\right)\ ,
\label{eq:short1}
\end{equation}
where $\langle \bar{\psi}\psi \rangle_0$ is the single flavour bilinear
quark condensate. $(\Pi_{\mathrm{VVP}})^{(abc)}_{\mu\nu}$
is an order parameter of the
spontaneous breaking of chiral symmetry. Hence, in the chiral limit,
it does not receive contributions from perturbative QCD at large
momentum transfers. This non-perturbative feature 
is in fact desirable to guarantee that the
OPE domain of applicability can be enlarged down to the 1-2 GeV energy region.
\par
In position space, Eq.~(\ref{eq:short1}) corresponds
to the limit where the space-time arguments of the three operators in
$\langle\mathrm{VVP}\rangle$ approach the same point at equal rates. We 
can also demand that only the argument of two of the three operators
converge towards the same point \cite{KN01}.
In this case two situations arise: 
either the two vector currents are taken at the same point, 
or one of the vector currents and the pseudoscalar density
are evaluated at the same argument.
The first situation 
was exploited in the analysis of the decay of pseudoscalars into 
lepton pairs of Ref.~\cite{KP99}. 
\par
We shall now build the $\langle\mathrm{VVP}\rangle$ Green's function
with the effective resonance theory given by $Z_{\mathrm{ERT}}[v,a,s,p]$, 
and impose that the short-distance constraint
in Eq.~(\ref{eq:short1}) is fulfilled. 

\begin{figure}[tb]
\begin{center}
\hspace*{-0.5cm}
\includegraphics[angle=0,width=0.8\textwidth]{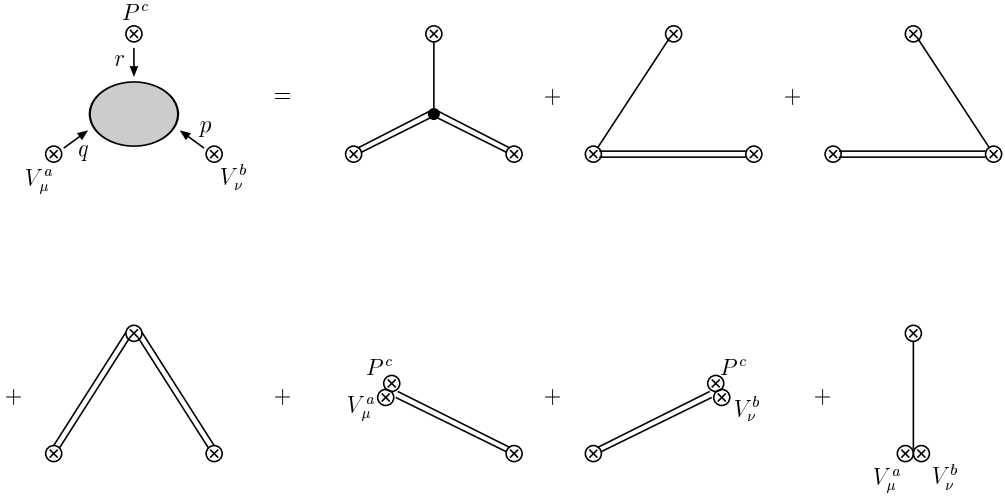}
\caption[]{\label{fig1} Diagrams entering the calculation of
the VVP 3-point function with the ERT action. Double lines
represent vector resonances, single lines are short for pseudoscalar
mesons.}
\end{center}
\end{figure}

At leading order in the $1/N_C$ expansion of QCD, the
three-point correlator is evaluated from the tree-level diagrams
shown in Fig.~\ref{fig1}. In this limit, an infinite spectrum of 
zero-width vector resonances should be considered in each channel. 
Fortunately, the LMD approximation to large-$N_C$, which assumes that 
a single resonance in each channel saturates the requirements of QCD,
can be invoked as a first test of the short-distance behaviour of
our Green's function. Indeed, we shall prove that
this approximation is sufficient to satisfy the short--distance
QCD constraints commented above.
\par
The couplings of the resonances among themselves and
to the external sources have been detailed in 
Eq.~(\ref{eq:allZ}), and the chiral limit is implied throughout. 
Our result reads
\begin{eqnarray}
\Pi_{\mathrm{VVP}}^{\mathrm{res}}(p^2,q^2,r^2)\!\!\!&=&\!\!\!
-\frac{\langle \bar{\psi}\psi \rangle_0}{F^2}
\,\Bigg\{\,
4\,F_V^2\,\frac{(d_1-d_3)\,r^2+d_3(p^2+q^2)}{(M_V^2-p^2)\,(M_V^2-q^2)\,r^2} 
\nonumber\\[3mm]
&&\!\!\!-2\sqrt{2}\,\frac{F_V}{M_V}\frac{r^2(c_1+c_2-c_5)+p^2(-c_1+c_2+c_5-2c_6)
+q^2(c_1-c_2+c_5)}{(M_V^2-p^2)\,r^2}
\nonumber\\[3mm]
&&\!\!\! -2\sqrt{2}\,\frac{F_V}{M_V}\frac{r^2(c_1+c_2-c_5)+
q^2(-c_1+c_2+c_5-2c_6)
+p^2(c_1-c_2+c_5)}{(M_V^2-q^2)\,r^2}
\nonumber\\[3mm]
&&\!\!\! +\frac{32F_V^2\,d_2}{(M_V^2-p^2)\,(M_V^2-q^2)}
\,-\,\frac{16\sqrt{2}F_V c_3}{M_V(M_V^2-p^2)}
\,-\,\frac{16\sqrt{2}F_V c_3}{M_V(M_V^2-q^2)}
\,-\,\frac{N_C}{8\pi^2r^2} 
\,\bigg\} \; .
\nonumber\\[3mm]
\label{eq:VVPres}
\end{eqnarray}
The contributions in Eq.~(\ref{eq:VVPres}) have been written 
following the same ordering of Fig.~\ref{fig1} (left to right, top to 
bottom). The last term originates from the
piece of the Wess-Zumino action $Z_{\mathrm{WZ}}[v,a]$
responsible of the pseudoscalar meson decays into two photons,
\begin{equation}
{\cal L}^{(4)}_{\mathrm{WZ}}=-\frac{\sqrt{2} \, N_C}{8\pi^2 \, F }
\ \epsilon_{\mu\nu\alpha\beta}
\,\langle \Phi \,\partial^{\mu}v^{\nu}\,
\partial^{\alpha}v^{\beta}\,
\rangle\,.
\label{eq:WZaction}
\end{equation}
\par
If we now take the limit of two momenta becoming simultaneously large in
$\Pi_{\mathrm{VVP}}^{\mathrm{res}}$, we find compatibility with the
QCD short-distance constraint up to
order $1/\lambda^4$, Eq.~(\ref{eq:short1}), provided the
following conditions among the ${\cal L}_V^{\mathrm{odd}}$ couplings hold:
\begin{eqnarray}
4 \, c_3+c_1&=&0 \; \; \; ,
\nonumber\\[3mm]
c_1-c_2+c_5 &=& 0 \; \; \; ,
\nonumber\\[3mm]
c_5-c_6&=& \frac{N_C}{64\pi^2}\frac{M_V}{\sqrt{2}F_V} \; \; \; ,
\nonumber\\[3mm]
d_1 + 8 \, d_2&=& -\frac{N_C}{64\pi^2}\frac{M_V^2}{F_V^2} \, + 
\, \frac{F^2}{4F_V^2}  \; \; \; ,
\nonumber\\[3mm]
d_3&=&-\frac{N_C}{64\pi^2}\frac{M_V^2}{F_V^2} \, + \, \frac{F^2}{8F_V^2}\,.
\label{eq:cond}
\end{eqnarray}
These relations have been obtained within the chiral limit. However
the couplings of the Effective Lagrangian do not depend on the masses
of the Goldstone fields and, consequently, the constraints in 
Eq.~(\ref{eq:cond}) apply for non--zero pseudoscalar masses too.
\par
Actually our $\langle\mathrm{VVP}\rangle$ three-point function
fully reproduces the LMD ansatz suggested in Ref.~\cite{M95}~:
\begin{equation}
\Pi_{\mathrm{VVP}}^{\mathrm{res}}(p^2,q^2,
(p+q)^2)= 
-\frac{\langle \bar{\psi}\psi \rangle_0 }{2}\, \cdot \, 
\frac{(p^2+q^2+r^2)-\frac{N_C}{4\pi^2}\frac{M_V^4}{F^2}}
{(p^2-M_V^2)(q^2-M_V^2)r^2}\, . 
\label{eq:VVPantsaz}
\end{equation}
As a consequence both the short--distance behaviour in Eq.~(\ref{eq:short1})
and those conditions where two vector currents or one vector current and
the pseudoscalar density meet at the same point, mentioned above, are
thoroughly satisfied.
\par
The ansatz (\ref{eq:VVPantsaz}) implies that 
we recover the LMD estimates for the low-energy 
constants derived in Ref.~\cite{KN01}.
The authors of this reference found that the same agreement with 
the short and long-distance QCD behaviour could not be reached
working with the resonance Lagrangian 
in the vector representation, not even at the expense of introducing local
contributions from the ${\cal O}(p^6)$ chiral Lagrangian.
They then suggested that the problem may be 
inherent to the effective Lagrangian approach and unlikely to be
fixed just by using other representations for the resonance fields; 
our result, derived in the antisymmetric tensor-field 
formulation with an odd-intrinsic-parity sector, contradicts this assertion, at
least in what concerns the $\langle\mathrm{VVP}\rangle$ Green's function. 
\par
Finally it is worth to comment the situation that would arise
if local ${\cal O}(p^6)$ operators of the chiral Lagrangian 
in Eq.~(\ref{eq:WZaction6}) were introduced in this analysis.
We argued in Section 2 that the couplings of those operators,
$t_i$, could be completely saturated by vector resonances and, accordingly, 
$t_i \simeq t_i^V$ and $\hat{t}_i \simeq 0$. If we include 
a non--vanishing $\hat{t}_i$ in the evaluation of the Green's function 
carried above it is easy to see that the high energy behaviour is 
spoiled unless higher--derivative couplings with resonances are 
considered. If we stay within our $Z_{\mathrm{ERT}}[v,a,s,p]$ action,
that satisfies by itself the matching with the QCD result,
the OPE imposes $\hat{t}_i = 0$, $i = 1,2$.
\par
It is also interesting to notice that the combinations of 
odd--intrinsic couplings which appear in the expressions of 
$t_i^V$, Eq.~(\ref{eq:t1t2}) are predicted from the QCD conditions
above. We obtain~:
\begin{eqnarray}
t_1^{\mathrm{V}} & = & \Frac{F^2}{64 \, M_V^4} \; \; , \nonumber \\
t_2^{\mathrm{V}} & = & - \, \Frac{N_C}{64 \, \pi^2 \, M_V^2} \, 
\left[ \, 1 \, - \, \Frac{4 \, \pi^2}{N_C} \, \Frac{F^2}{M_V^2} \, \right]
\; \; ,
\label{eq:t1t2ex}
\end{eqnarray}
which coincide~\footnote{There is a minus sign difference in the 
definitions of $t_1$ and $t_2$ in \cite{M95} because 
the convention used there for the Levi-Civita 
tensor is the opposite to ours.} with the predictions made for 
these parameters in
\cite{M95}. This fact is not surprising, since
the relations (\ref{eq:t1t2ex}) were derived in \cite{M95} by expanding 
the $\langle\mathrm{VVP}\rangle$ ansatz, Eq.~(\ref{eq:VVPantsaz}), 
at low-momenta and comparing it with the $\langle\mathrm{VVP}\rangle$ 
expression obtained from the ${\cal L}^{(6)}_{\mathrm{odd}}$ Lagrangian. 
The success in reproducing the same representation
for $\Pi_{\mathrm{VVP}}$ within the resonance effective Lagrangian
has automatically generated
identical values for the chiral parameters~\footnote{The
author of \cite{M95} extended the results for $t_1^V$ and $t_2^V$ above by
including an additional pole-contribution in the
VVP ansatz from a pseudoscalar $\pi(1300)$ resonance.}.
   
\section{Phenomenology of intrinsic-parity violating processes}
\label{sec:pheno}
\hspace*{0.5cm}
Odd--intrinsic--parity processes have been widely studied within 
chiral perturbation
theory where resonances are integrated out \cite{anBij}.
In order to gain more insight on the odd-intrinsic-parity sector of the  
resonance Lagrangian and, to make some test on
the validity of the short-distance conditions obtained above, we study 
in this section the 
processes $\omega\to\pi\gamma$, $\omega\to3\pi$ and $\pi\to 2\gamma$.

\subsection{$\omega\to\pi\gamma$ }
\begin{figure}[tb]
\begin{center}
\hspace*{-0.5cm}
\includegraphics[angle=0,width=0.8\textwidth]{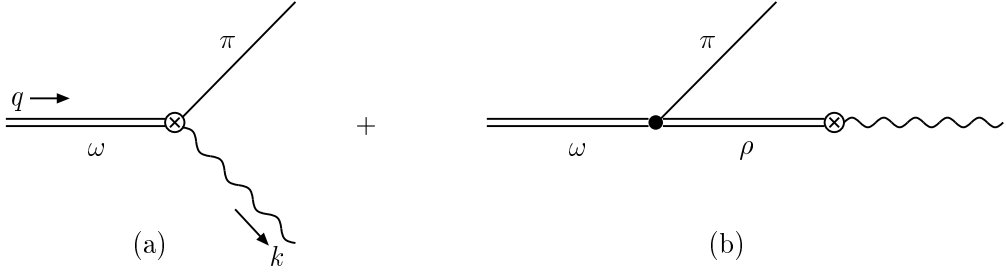}
\caption[]{\label{fig2}  Lowest order diagrams for the process
$\omega\to\pi\gamma$.}
\end{center}
\end{figure}

\hspace*{0.5cm}
At tree-level, the intrinsic-parity violating transition $\omega\to\pi\gamma$ 
receives contributions from
both the VJP and VVP terms of ${\cal L}_V^{\mathrm{odd}}$. 
The corresponding diagrams are displayed in Fig.~\ref{fig2}. 
The physical $\omega$ resonance is a superposition of an
octet component, $\omega_8$,
and a singlet one, $\omega_1$, which can be added as a diagonal 
matrix $\omega_1/\sqrt{3}$ to the octet, Eq.~(\ref{eq:vmultiplet}); if 
ideal mixing is assumed then the states of defined mass are
$$
|\, \omega \, \rangle = \sqrt{\,\frac{2}{3}}\,| \,\omega_1 \rangle+
\sqrt{\,\frac{1}{3}}\,|\,\omega_8\rangle  \; \; ,
$$
and
$$
|\, \phi \, \rangle = -\sqrt{\,\frac{1}{3}}\,| \,\omega_1 \rangle+
\sqrt{\,\frac{2}{3}}\,|\,\omega_8\rangle \,.
$$

The amplitudes for the direct and $\rho$-mediated diagrams, 
Figs.~\ref{fig2}a and \ref{fig2}b respectively, read
\begin{eqnarray}
i\,{\cal M}^{direct}_{\omega\to\pi\gamma} \!\!\! &=& \!\!\!
i\,\epsilon_{\alpha\beta\rho\sigma}
\,\epsilon^{\alpha}_{\omega}\epsilon^{\beta}_{\gamma}q^{\rho}k^{\sigma}
\,2\sqrt{2}\,\frac{e}{M_{\omega}M_V F}\,
\bigg[ \,(c_2-c_1+c_5-2c_6)\,M_{\omega}^2+
(c_1+c_2+8c_3-c_5)\,m_{\pi}^2 \,\bigg] \; ,
\nonumber\\[3mm]
i\,{\cal M}_{\omega\to\pi\gamma}^{\rho} \!\!\! &=& \!\!\!
-i\,\epsilon_{\alpha\beta\rho\sigma}
\,\epsilon^{\alpha}_{\omega}\epsilon^{\beta}_{\gamma}q^{\rho}k^{\sigma}
\frac{4\,e}{M_{V}^2 M_{\omega}}\,\frac{F_V}{F}\,\bigg[ \,d_3 \,M_{\omega}^2
+(d_1+8d_2-d_3)m_{\pi}^2 \,\bigg]\,,
\label{eq:M-w-pi-ga}
\end{eqnarray}
where we have kept the generic mass $M_V$ of the meson octet in the
$\rho$ propagator, in consistency
with the procedure followed in the analysis of Section \ref{sec:short};
distinction is made between $M_V$ and $M_{\omega}$ when the latter is
of kinematic origin. 
Quite remarkably, if we now plug in the QCD constraints, Eq.~(\ref{eq:cond}),
obtained from the analysis of the short-distance behaviour of the 
$\langle\mathrm{VVP}\rangle$ Green's function, we find a full prediction for 
this process~:
\begin{equation}
i\,{\cal M}_{\omega\to\pi\gamma}=i\,\epsilon_{\alpha\beta\rho\sigma}
\,\epsilon^{\alpha}_{\omega}\epsilon^{\beta}_{\gamma}q^{\rho}k^{\sigma}
\frac{e}{F_V}\,\bigg[ \,\frac{N_C}{8\pi^2}\,\frac{M_{\omega}}{F}
\,-\frac{F}{2}\,\frac{M_{\omega}}{M_{V}^2} \, \left( 1 + 
\Frac{m_{\pi}^2}{M_\omega^2} \, \right)\,\bigg]\,.
\label{eq:w-pi-ga}
\end{equation}
We notice that the direct (Fig.~\ref{fig2}a) and the $\rho$ exchange
diagrams (Fig.\ref{fig2}b) almost contribute to similar extent 
to this process. This means that contrary to what we would expect from 
vector meson dominance, the $\omega\rho\pi$ coupling does not saturate  
the decay $\omega\to\pi\gamma$. The actual value
of this coupling in our formalism~\footnote{The tiny contribution coming
from the pion mass contribution in ${\cal M}^{\rho}_{\omega \to \pi \gamma}$
can be obviated in this discussion.}, $d_3$, is less than half 
of the one that would arise from VMD, where only the diagram 
Fig.~\ref{fig2}b contributes. This has immediate consequences
to other decay channels, as we shall see in the next subsection. 
\par
Finally, the width is easily obtained, giving
\begin{equation}
\Gamma(\omega\to\pi\gamma)=\frac{\alpha}{192}\,M_{\omega}
\bigg( 1-\frac{m^2_{\pi}}{M_{\omega}^2}\bigg)^3\,
\bigg[\,\frac{N_C}{4\pi^2}\,\frac{M_{\omega}^2}{F^2}
\,-\frac{M_{\omega}^2}{M_{V}^2} \left( 1 + 
\Frac{m_{\pi}^2}{M_\omega^2} \, \right)
\,\bigg]^2\,.
\label{eq:Gamma_w}
\end{equation}
The relation $F_V=\sqrt{2} \, F$, consequence of conditions (\ref{eq:FVGV}) 
and (\ref{eq:FVGVa}), has been employed
in deriving the result in Eq.~(\ref{eq:Gamma_w}).   
Varying the parameter $F$ from the bare value
$F_0\simeq 87$ MeV to the dressed one (i.e. the pion decay constant),
$F_{\pi} \simeq 92.4$ MeV \cite{PDG}, we get that 
our prediction for $\Gamma(\omega\to\pi\gamma)$ ranges
from 0.703 MeV to 0.524 MeV, with the choices $M_V=M_{\rho}=771.1$ MeV and
$M_{\omega}=782.6$ MeV \cite{PDG}. 
This 5--30\% deviation from
the experimental value, 
$\Gamma(\omega\to\pi\gamma)|_{\mathrm{exp}}=(0.734\pm 0.035)\,\mathrm{MeV}$,  
is in accordance with the expected size of next-to-leading $1/N_C$ corrections. 
\par
Our result for $\Gamma ( \omega \to \pi \gamma )$ is quite significant being
a pure prediction of the matching procedure of the resonance effective
theory with the OPE expansion given by QCD. The extension of our analysis
 to other decay channels (e.g.
$K^*\to K\gamma$, $\phi\to\eta\gamma$)
requires that exact SU(3)$_{\mathrm{V}}$ symmetry is left aside
in order not to lose the predictive power shown in $\omega \to \pi \gamma$. 
This study would require to consider the OPE expansion
in the asymptotic regime
keeping distinct masses for each quark flavour, a rather non trivial
task.

\subsection{$\omega\to\pi^+\pi^-\pi^0$}
\begin{figure}[tb]
\begin{center}
\hspace*{-0.5cm}
\includegraphics[angle=0,width=0.80\textwidth]{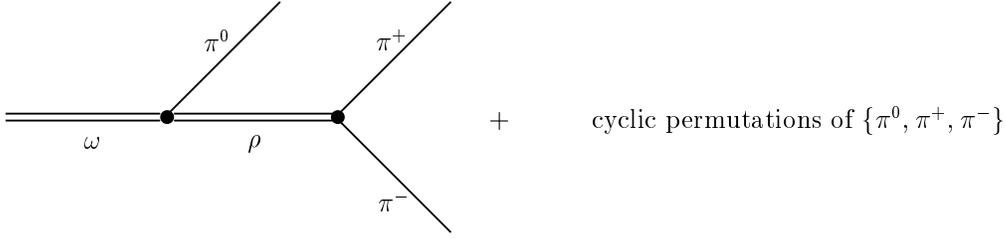}
\caption[]{\label{fig3} The $\omega\to\pi^+\pi^-\pi^0$ decay amplitude
via an intermediate $\rho$ exchange.}
\end{center}
\end{figure}

\hspace*{0.5cm} 
The odd-intrinsic parity sector included in the resonance Lagrangian can also 
account for the $\rho$-mediated mechanism of decay of the
$\omega$ meson to the $\pi^+\pi^-\pi^0$ final state, Fig.~\ref{fig3}. If we
label as $k_1,\,k_2,\,k_3$ the momenta of the $\pi^+\,,\pi^-$ and $\pi^0$
respectively, the amplitude associated to the diagram of Fig.~\ref{fig3},
including cyclic permutations among $k_1,\,k_2$ and $k_3$, reads

\begin{eqnarray}
i\,{\cal M}_{\omega\to3\pi}=i\,\epsilon_{\alpha\beta\rho\sigma}
\,k_1^{\alpha}k_2^{\beta}k_3^{\rho}\epsilon^{\sigma}_{\omega}
\frac{8\,G_V}{M_{\omega}F^3}\,\Bigg[&&\!\!\!\!\!\!\!\!
\frac{m^2_{\pi}(d_1+8d_2-d_3)
+(M_{\omega}^2+s_{12}) \, d_3}{M_V^2-s_{12}}
\nonumber\\[3mm]
&&\!\!\!\!\!\!\!\!+\,\{s_{12}\to s_{13}\}+\{s_{12}\to s_{23}\}\,
\Bigg]
\,.
\label{eq:w-3pi}
\end{eqnarray}
The kinematic invariants are defined as usual, i.e. $s_{ij}=(k_i+k_j)^2$.
The VMD hypothesis for this decay predicts that the amplitude above is 
the dominant one. Then the corresponding width would be calculated as
\begin{eqnarray}
\label{eq:Gamma_w3pi} 
\Gamma(\omega\to \pi^+ \pi^- \pi^0) & = & 
\frac{G_V^2}{4 \, \pi^3 \, M_{\omega}^5 \, F^6}\,
\, \int_{4m_{\pi}^2}^{(M_\omega-m_{\pi})^2}ds_{13}
\int_{s_{23}^{min}}^{s_{23}^{max}}ds_{23}\,{\cal P}(s_{13},s_{23}) \, \times
\\[3mm]
&& \times\Bigg[  \, \Frac{m^2_{\pi}(d_1+8d_2-d_3)
+(M_{\omega}^2+s_{12}) \, d_3}{M_V^2-s_{12}}
+\,\{s_{12}\to s_{13}\}+\{s_{12}\to s_{23}\}
\, 
\Bigg]^2\,, \nonumber 
\end{eqnarray}
where the function ${\cal P}$ is the polarization average of the
tensor structure of ${\cal M}_{\omega\to3\pi}$, 
\begin{equation}
{\cal P}(s_{13},s_{23})=\frac{1}{12}\left\{ -m_{\pi}^2(m_{\pi}^2-
M_{\omega}^2)^2-
s_{13}s_{23}^2+(3m_{\pi}^2+M_{\omega}^2-s_{13})s_{13}s_{23}\right\}
\,.
\end{equation} 
With $G_V = F / \sqrt{2}$ and the relations obtained by the 
short-distance matching, we find that the width above works out 
$\Gamma (\omega \rightarrow \pi^+ \pi^- \pi^0) \simeq 1.4 \, \mathrm{MeV}$, 
quite far from the experimental result \cite{PDG}, 
$\Gamma(\omega\to \pi^+ \pi^- \pi^0)|_{\mathrm{exp}}= (7.52\, \pm \, 0.06) \,
\mathrm{MeV}$. 
Clearly, the contribution from a direct $\omega\to 3\pi$ amplitude
must be larger than expected from VMD. 
Such deviation can be traced back to the result obtained
in the previous section for $\omega\to\pi\gamma$. There we
found that the $d_3$ parameter was less than half the value one should 
expect from a dominant role of the $\rho\omega\pi$ coupling. 
The $\omega\to 3\pi$ width calculated above, Eq.~(\ref{eq:Gamma_w3pi}),
is essentially (neglecting the tiny piece driven by the pion 
mass squared) proportional to $d_3^2$; therefore, there is 
roughly a factor of $\sim 4$ 
between our calculation of $\Gamma(\omega\to\pi\rho \to 3\pi)$ and the
result obtained under VMD by fixing the $\rho\omega\pi$ coupling 
from the $\omega\to\pi\gamma$ width 
(see for example Ref.~\cite{EU02}). 
This factor would raise the result of (\ref{eq:Gamma_w3pi})
to $\sim 5.6 \, \mathrm{MeV}$, i.e.  
reaching the level of accuracy 
of leading large-$N_C$ calculations.
\par
According to the precedent discussion, 
the intermediate meson exchange does not account entirely for 
the $\omega$ decay into three pions, and the direct terms must be 
considered~\footnote{In our effective theory, these
terms would be obtained by 
writing down the operators which give rise to local contributions to
$\omega\to 3\pi$.}. In fact both contributions appear at the same
order in the large--$N_C$ expansion and 
the $\rho$ resonance, being
far off--shell in this process, does not resonate.
Consequently, there is no reason that justifies
neglecting the direct vertex. 
Indeed, it was pointed out in Ref.~\cite{EU02}
that VMD alone predicts a too large $\rho\omega\pi$ coupling with
respect to what suggests naive chiral counting. 
The QCD--enforced appearance of a direct term 
in our approach, which has reduced 
the $\rho\omega\pi$ coupling to the half, 
casts some light on the issue. 

\subsection{$\pi\to\gamma\gamma$ }
\hspace*{0.5cm} 
In the chiral limit, the amplitude for the $\pi\to\gamma\gamma$ process
is non-vanishing and exactly predicted by
the ABJ anomaly \cite{ABJ69}, Eq.~(\ref{eq:WZaction}). Away from this limit, 
the amplitude receives small contributions from different sources, 
including isospin-breaking effects, as well as 
electromagnetic and higher-order chiral corrections. As the loop contribution
vanishes \cite{DH85},
the latter corrections start with the ${\cal O}(p^6)$ Goldstone
chiral Lagrangian.
The odd-intrinsic-parity interactions among vector resonances
introduced in Section \ref{sec:RChPT} also generate 
chiral corrections to this
process proportional to $m_{\pi}^2$. Let us first 
study the numerical size of these corrections, fixed by virtue of the
short-distance constraints. 

\begin{figure}[tb]
\begin{center}
\hspace*{-0.5cm}
\includegraphics[angle=0,width=0.7\textwidth]{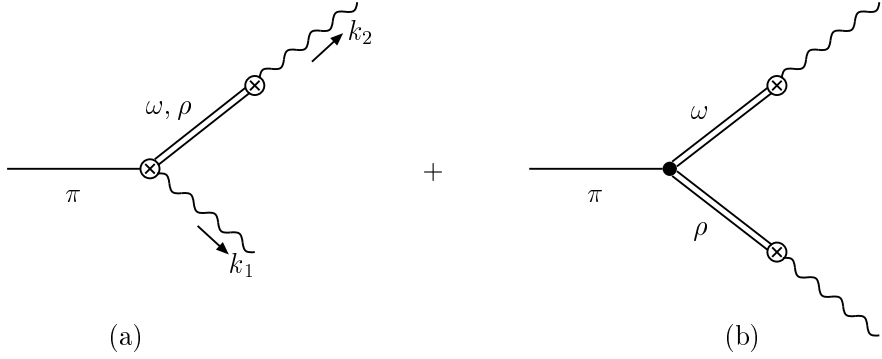}
\caption[]{\label{fig4} Feynman diagrams with vector mesons giving
${\cal O}(m_{\pi}^2)$ corrections to $\pi\to\gamma\gamma$ decay. The
diagrams with reversed photon momenta must be added.}
\end{center}
\end{figure}

The amplitudes for the decay via
intermediate meson exchange, depicted in Fig.~\ref{fig4}, give as a result 
\begin{eqnarray}
i\,{\cal M}^{(a)}_{\pi\to\gamma\gamma}&=&-i\,\epsilon_{\alpha\beta\rho\sigma}
\,\epsilon^{\alpha}_{1}\epsilon^{\beta}_{2}k_1^{\rho}k_2^{\sigma} \;
\,\Frac{8\sqrt{2}}{3}\,\Frac{e^2}{M_V} \, \Frac{F_V}{F} \, 
\, \frac{m^2_{\pi}}{M_{V}^2}\,
\Big( c_1+c_2+8c_3-c_5 \Big)\, \nonumber \\[4mm]
& = & \; 0 \; \; \; ,\nonumber\\[8mm]
i\,{\cal M}^{(b)}_{\pi\to\gamma\gamma}&=&i\,\epsilon_{\alpha\beta\rho\sigma}
\,\epsilon^{\alpha}_{1}\epsilon^{\beta}_{2}k_1^{\rho}k_2^{\sigma}
\, \; 
\frac{8 \, e^2}{3\, F}\,\frac{F_V^2}{M_{V}^2} \, 
\frac{m^2_{\pi}}{M_{V}^2} \,
\Big( d_1+8d_2-d_3\Big)
\nonumber\\[3mm]
&=&i\,\epsilon_{\alpha\beta\rho\sigma}
\,\epsilon^{\alpha}_{1}\epsilon^{\beta}_{2}k_1^{\rho}k_2^{\sigma}
\,\Frac{e^2 \, F}{3M_V^2} \,\Frac{m^2_{\pi}}{M_{V}^2} \, 
\,.
\label{eq:pi-ga-gaV}
\end{eqnarray}
The diagram with a VJP vertex vanishes after the short-distance conditions
are applied, and the remaining contribution gets completely fixed. 
The correction induced into the $\pi \to \gamma \gamma$ width, by
our result above gives~:
\begin{equation}
\Gamma(\pi\to\gamma\gamma)=\frac{\alpha^2}{64 \, \pi^3 \, F^2}m_{\pi}^3
\left[ \, 1 \, - \, \Delta \, \right]^2 \; \; ,
\end{equation}
where
\begin{equation}
\Delta \; = \; 
\Frac{4\pi^2}{3} \, \Frac{F^2}{M_V^2} \; 
\Frac{m_{\pi}^2}{M_{V}^2} \,\, \simeq \; \; 0.006 \; \; \; .
\end{equation}
This result provides a tiny 1$\%$ correction to the width, and it is
perfectly compatible with the experimental uncertainty, 
$\Gamma(\pi\to\gamma\gamma)|_{\mathrm{exp}}=(7.7\pm 0.6)$ eV.
\par
This evaluation of the amplitude for the $\pi \to \gamma \gamma$ process
could also have been carried out within the chiral Lagrangian 
${\cal L}_{\mathrm{odd}}^{(6)}$ of Eq.~(\ref{eq:WZaction6}), where only the
operator with $t_1$ contributes~\footnote{The operator with the $t_2$ 
coupling only contributes  if one of the photons is off--shell.}. With
$t_1 \simeq t_1^V$ and using the value given in Eq.~(\ref{eq:t1t2ex}) 
we obtain the result above. The exercise carried out in this Subsection,
evaluating the diagrams in Fig.~\ref{fig4}, shows explicitly that 
only the two--resonance driven amplitude gives contribution to this 
process in the antisymmetric formulation.

\section{Conclusions}
\hspace*{0.5cm}
Effective theories of QCD carry all--important features of the 
underlying theory to describe the relevant hadron dynamics in the
non--perturbative regime. The odd--intrinsic--parity 
sector has been studied within chiral perturbation
theory but its extension to the energy region of the resonances
requires a proper implementation of the active degrees of freedom and
to generate the effective theory through a procedure able to 
enforce the relevant dynamics on the coupling constants. This task 
has been addressed in this paper.
After considering the operators of the Lagrangian, that rely 
on the global symmetries of QCD, 
we proceed to drive the information, from the underlying
theory onto the couplings, through a matching 
with the leading OPE of the Green's function in the chiral limit.
\par
Let us highlight the main results that can be extracted from the
previous sections. 
\begin{itemize}
\item[$\bullet$] The lowest order Lagrangian involving interactions
among one Goldstone mode and two vector particles has been introduced 
in the Resonance Chiral Theory with the vector resonances 
described in terms of antisymmetric tensor fields.
\item[$\bullet$] The vector-vector-pseudoscalar three-point-function
$\langle\mathrm{VVP}\rangle$
has been calculated at tree level with the new sector added to 
the resonance Lagrangian. Assuming that a matching procedure
between the result obtained from the effective action and 
from QCD in terms of massless quarks is reliable at large momenta, 
we have derived a set of relations
among the parameters of the odd-intrinsic-parity sector.
\item[$\bullet$] In contrast to the result
of Ref.~\cite{KN01}, where vector resonances were described in the
Proca formalism, the expression for the 
$\langle\mathrm{VVP}\rangle$ Green's function obtained from
the Lagrangian with antisymmetric tensor fields is fully compatible
with the short--distance QCD constraints, which reduce it 
to the ansatz suggested by LMD in the large-$N_C$ limit of QCD,
successfully tested in previous works \cite{M95,KP99}. 
\item[$\bullet$] On the way, we have found that the same combinations
of couplings which appears in the short-distance
QCD constraints, show up in the $\omega\to\pi\gamma$ 
amplitude calculated with the resonance Lagrangian, thus allowing
us to give a full prediction for this decay. The
agreement with the experimental value is remarkable. 
\item[$\bullet$] The $\omega\to\pi\gamma$ calculation above shows an
important feature: the contribution from a direct $\omega\pi\gamma$ 
vertex is larger than expected from VMD. 
Indeed, it amounts to more than 50$\%$
of the total result for this amplitude. This agrees with the 
expectations from the $1/N_C$ counting, as both mechanisms contribute
to the same order.
\item[$\bullet$] The last point has an important consequence
for other channels where VMD alone was thought to be the
relevant mechanism of decay.  
To serve as an example, we have shown 
that the intermediate meson exchange $\omega\to\rho\pi\to 3\pi$ 
cannot dominate the $\omega\to 3\pi$ process in our framework, 
and the local contribution thus becomes essential. 
\end{itemize}
Our study has shown that the use of effective theories of QCD in the
intermediate energy region, populated by resonances, 
endows the basic information to provide both qualitative
and quantitative descriptions of the hadron phenomenology in a 
model--independent way. Consequently it provides a compelling 
framework to work with.

\vspace*{1cm} 
\noindent
{\large \bf Acknowledgements}\par
\vspace{0.2cm}
\noindent 
We wish to thank V. Cirigliano for his helpful comments and for reading
the manuscript.
The work of P.~D. Ruiz-Femen\'\i a has been partially supported by a FPU
scholarship of the Spanish {\it Ministerio de Educaci\'on y Cultura}. 
J. Portol\'es is supported by a \lq \lq Ram\'on y Cajal" contract with CSIC
funded by MCYT.
This work has been supported in part by TMR EURIDICE, EC Contract No. 
HPRN-CT-2002-00311, by MCYT (Spain) under grant FPA2001-3031, and
by ERDF funds from the European Commission.

\appendix
\newcounter{erasmo}
\renewcommand{\thesection}{\Alph{erasmo}}
\renewcommand{\theequation}{\Alph{erasmo}.\arabic{equation}}
\renewcommand{\thetable}{\Alph{erasmo}}
\setcounter{erasmo}{1}
\setcounter{equation}{0}
\setcounter{table}{0}

\section*{Appendix  }
\hspace*{0.5cm}
Within the antisymmetric formulation, the integration 
of a vector meson gives a contribution which starts at
${\cal O}(p^4)$ in the chiral counting.
Interaction terms
with a Levi-Civita tensor 
start to contribute~\footnote{For terms involving 
vector resonances, this counting should be understood as
the one obtained after integrating out the resonances, i.e. the order
of the chiral operator induced by vector exchange.} at ${\cal O}(p^6)$, 
as terms
with one vector meson and an ${\cal O}(p^2)$ chiral tensor are not
charge conjugation or parity invariant, and a possible term with 
two resonance fields,
$\epsilon_{\mu\nu\rho\sigma}\langle V^{\mu\nu}V^{\rho\sigma}\rangle$,
is forbidden by parity conservation. Besides, terms of odd order, i.e. 
${\cal O}(p^3)$ or ${\cal O}(p^5)$, cannot be written down in the presence
of an $\epsilon_{\mu\nu\rho\sigma}$ tensor.
Either a chiral tensor
of ${\cal O}(p^4)$ together with a vector meson is needed, giving rise to the
VJP terms, or two vector resonances and a 
chiral tensor of ${\cal O}(p^2)$ (VVP terms).  
\par
The available chiral tensors have already been introduced
in Section~\ref{sec:RChPT}: $\chi_{\pm}$, $f_{\pm}^{\mu\nu}$ are ${\cal O}(p^2)$,
while the covariant derivative $\nabla_{\alpha}$ and $u_{\alpha}$
count as ${\cal O}(p)$. These tensors have defined 
transformation properties under chiral rotations and thus allow us
to write down chiral invariant objects in a straightforward way. 
\par
Let us first give some clues about the construction of the VVP basis. 
Aside
from the two vector mesons, we should consider
all possible tensors giving one pseudoscalar. Therefore, we can 
have: 
\begin{itemize}
\item One covariant derivative $\nabla_{\mu}$ and one $u_{\nu}$
tensor, with the covariant derivative acting on either the resonance fields
or the pseudoscalar $u_{\nu}$. In the latter case
$\nabla_{\mu}u_{\nu}$ is symmetric in its indices for the linear term
of the expansion of $u_{\nu}$ in terms of Goldstone fields:
$$
u_{\nu}=-\frac{\sqrt{2}}{F}\,\partial_{\nu}\Phi+ 
\mathrm{terms\ with\ 3\ pseudoscalar\ fields} + \dots \; \; .
$$
\item A $\chi_{-}$ external field, whose expansion in terms of
the pseudoscalar octet of fields starts with one particle states. A 
$\chi_{+}$ external field together with the two vector mesons is
however not allowed by parity conservation. 
\end{itemize}
In addition, the Schouten identity, 
\begin{equation}
g_{\rho\sigma}\epsilon_{\alpha\beta\mu\nu}+
g_{\rho\alpha}\epsilon_{\beta\mu\nu\sigma}+
g_{\rho\beta}\epsilon_{\mu\nu\sigma\alpha}+
g_{\rho\mu}\epsilon_{\nu\sigma\alpha\beta}+
g_{\rho\nu}\epsilon_{\sigma\alpha\beta\mu}=0\,,
\label{eq:Schouten}
\end{equation}
reduces the number of independent operators because it may establish
relations among those with different ordering of the Lorentz indices. 
As an example, consider the two following VVP terms:
\begin{eqnarray}
{\cal O}^1&=&\epsilon_{\mu\nu\rho\sigma}\,
\langle \, \{V^{\mu\nu},V^{\rho\alpha}\} \nabla_{\alpha}u^{\sigma}\,\rangle
=g_{\alpha\lambda}\,\epsilon_{\mu\nu\rho\sigma}\,
\langle \, \{V^{\mu\nu},V^{\rho\lambda}\} \nabla^{\alpha}u^{\sigma}\,\rangle\,,
\nonumber\\[3mm]
{\cal O}^2&=&\epsilon_{\mu\nu\rho\sigma}\,
\langle \, \{V^{\mu\nu},V^{\rho\sigma}\} \nabla_{\alpha}u^{\alpha}\,\rangle
=g_{\alpha\sigma}\,\epsilon_{\mu\nu\rho\lambda}\,
\langle \, \{V^{\mu\nu},V^{\rho\lambda}\} \nabla^{\alpha}u^{\sigma}\,\rangle\,.
\nonumber
\end{eqnarray}
With the identity (\ref{eq:Schouten}) we find that the second operator is 
proportional to the first one:
$$
{\cal O}^2=4\,{\cal O}^1\,.
$$
Similarly, the Schouten identity must be applied 
to operators with the $\nabla_{\mu}$ 
acting on the resonance fields and to operators from the VJP sector
to further reduce the basis. 
\par
To close with the analysis of the VVP interactions,
recall that a term 
$\sim \langle V^{\mu\nu}V^{\rho\sigma}f_{-}^{\alpha\beta}\rangle$ 
would include an external vector (or axial-vector) 
source in addition to the wanted pseudoscalar. Clearly, these terms do 
not belong to our VVP sector. 
\par
For the VJP interactions, basically the same considerations made above hold,
and the substitution of one of the resonance fields $V^{\mu\nu}$ by
and external vector field $f_{+}^{\mu\nu}$, which has the same
properties under $P$ and $C$ transformations, gives the allowed
VJP structures. Note that for each VVP term
two VJP operators emerge with this procedure (except for
the term with $\chi_-$), as the vector
tensors are not equal now. We have chosen that 
$\nabla^{\alpha}$ acts on the vector meson or on the pseudoscalar
field
to define the final set of independent VJP operators. 
As quoted in the main text, the term 
${\cal O}_{\mbox{\tiny VJP}}^4$, where the
pseudoscalar now comes up from the $f_{+}^{\mu\nu}$
tensor, is a SU(3)$_\mathrm{V}$-breaking operator.
Indeed its lower order expansion in terms of Goldstone fields is
proportional to $m_{K}^2-m_{\pi}^2$.

\end{document}